# Analyzing Potential Solutions Involving Regulation to Escape Some of AI's Ethical Concerns

Jay Nemec

## Abstract

Artificial intelligence (AI), although not able to currently capture the many complexities of humans, are slowly adapting to have certain capabilities of humans, many of which can revolutionize our world. AI systems, such as ChatGPT and others utilized within various industries for specific processes, have been transforming rapidly. However, this transformation can occur in an extremely concerning way if certain measures are not taken. This article touches on some of the current issues within the artificial intelligence ethical crisis, such as the concerns of discrimination within AI and false information that is becoming readily available with AI. Within this article, plausible solutions involving regulation are discussed and how they would mitigate ethical concerns. These include the self-regulation of businesses along with government regulation, and the effects these possible solutions can both have on current AI concerns.

## Introduction

From the entertainment industry to the healthcare industry, artificial intelligence (AI) is expected to play an impact on many various fields in our future. This tool has many benefits, ranging from being able to chat with healthcare patients about what may be the problem in their body (Moor, et al., 2023) to generating new types of art and media from a simple text entry box (Davenport, Mittel, 2022). However, these benefits do not come without their drawbacks. Worries about utilizing AI's capabilities for unfair purposes may be a key concern and is a pivotal reason why businesses, governments, organizations, and others that utilize AI or are planning to, should be cautious. Ethical concerns such as discrimination, false information being generated, and businesses having too much power, are some of the major problems within the world of innovation today, specifically in this diverse AI field. This is a problem that is important to be dealt with by government representatives and businesses. Government representatives are the ones with the ability to instill government regulations, while businesses are the ones that can instill personal regulations on themselves.  Having businesses and government representatives understand these key issues and instilling respective regulations could prove to be a pivotal first step to solving some of AI's ethical concerns.

**Various AI Ethical Concerns**

**i) Discrimination and Bias within AI Systems**

AI's ethical concerns bring up some issues such as discrimination that was thought to be left in the past. An article from the technology magazine, Wired, addresses how discrimination with AI is an extreme problem within the hiring process. The author states, "But in practice it (AI) has sometimes done the opposite, leading the US government to warn employers about the potential for algorithms to discriminate against people with disabilities" (Knight, 2019, para. 6). This emphasizes one worry of utilizing AI in areas such as job recruiting, and how these systems can hurt many people's chances at being able to fairly get jobs. Even though these systems have the ability to improve workplace diversity if utilized correctly, they are being utilized in the opposite fashion.

Additionally, discrimination may be a significant issue if AI is utilized in certain areas such as within the legal system. For example, in Broward County, Florida, a privately owned AI system was put into effect to attempt to solve the problem of overcrowding within jails. This system gave a score based on how likely repeated criminal activity was expected from certain individuals. However, the AI system utilized was seemingly defective as many of the scores predicted by the system were inaccurate (Mattu et al., 2016). This is shown within ProPublica's article, as the author states, "The score proved remarkably unreliable in forecasting violent crime: Only 20 percent of the people predicted to commit violent crimes actually went on to do so" (Mattu et al., 2016, Two Petty Theft Arrests section, para. 4). An unreliable system such as this one should not be utilized in the United States, especially not for a task that is determining how people's futures will turn out. Moreover, the article stated that black defendants were expected to commit a crime twice as many times as white defendants, to which many black defendants were wrongly labeled (Mattu et al., 2016). This is extremely concerning and shows another example of essentially dictating a person's life with a possibly faulty system. This is one important example of where government or business self regulation is necessary to help protect others, as this can help balance risks that a system like this causes.

Bias can also sometimes clearly be shown when an AI system is used. For example, I used two different free AI image systems to test out a possible area of bias that AI might possess. These systems were Nvidia's Stable Diffusion XL model and Hotpot's AI image generator. When the words "female basketball player" were entered into these three system's inputs, both systems outputted an image of a colored woman. This clearly represents the underlying bias that occurs within some of these system's programming. With Hotpot's image generator, a white woman was not shown until the word "white" was added into the input. Nvidia's system was much more accurate, showing a white player the second time with no additional input needed. This emphasizes a major concern involving how certain AI systems are programmed, as discrimination could easily be witnessed if a system is not programmed correctly.

This bias is discussed within an article by UNESCO, which is an organization focused on education that was initiated by the United Nations. Within the article, the author states, "An image search for "school girl" will most probably reveal a page filled with women and girls in all sorts of sexualised costumes. Surprisingly, if you type "school boy", results will mostly show ordinary young school boys. No men in sexualised costumes or very few" (UNESCO, 2023, Biased AI section, para. 2). This once again shows an extremely common drawback of AI systems and shows how if bias is inputted (by accident or on purpose) during a system's learning period, the system will input these biases into its prior knowledge. Then, when someone asks a similar or the same question to the system, it may produce discriminatory or biased responses.

**ii) False Information**

Posting fake footage and information is another major ethical challenge that AI causes, and is a key concern that needs to be addressed when considering the overall implications of AI on society. With all the concern about real data being leaked, the fact that false information could end up online should not surprise anyone. An article from the technology website Wired explains more about the major impact of AI on today's society. It states, "Deepfakes, for example---AI-generated videos meant to look like real footage---are now accessible to anyone with a laptop" (Barber, 2019, para. 5). This proves that fake videos among other incorrect information could easily end up being posted on the Internet due to AI's new generative nature, with people thinking that these events actually occurred.

Additionally, many people are concerned as ChatGPT and other AI generative softwares are able to make up their own false evidence to certain questions, which is a key reason why it should not be utilized as a reliable source. The University of Waterloo supports this claim in one of their articles about generative AI and they explain how important it is to be fact checking the information that comes out of these systems (UWaterloo, 2023). They additionally emphasize the importance of fact-checking information if it could have a negative impact on someone and their reputation. For example, ChatGPT and other AI softwares are known to sometimes make false criminal accusations on people that are untrue. A recent case involves AI falsely accusing a professor at George Washington University of sexual harassment (Turley, 2023). When generative AI starts making false claims such as this one that can stretch into a criminal realm, there is no other option than having some sort of regulation upon it.

## A Possible Solution to These Concerns With Business Self-Regulation

A plausible solution to some of the concerns within this AI ethical dilemma would be businesses agreeing to regulate themselves. With key issues such as discrimination and false information becoming much more common in today's world, these businesses will need to apply several key steps in the path to successful self-regulation.

The businesses will first need to analyze the exact background of how their system was trained, as AI systems are simply built off training data. An article from the Tableau database describes this process: "At the core level, an AI algorithm takes in training data (labeled or unlabeled, supplied by developers, or acquired by the program itself) and uses that information to learn and grow. Then it completes its tasks, using the training data as a basis" (*Artificial Intelligence (AI) Algorithms: A Complete Overview*, n.d., How do AI algorithms work? section, para. 2). This quote shows that the base data needs to be monitored. An AI system will never be more or less biased than its training data and the person who programs it. Therefore, it is essential that the programmer helping train the AI system is not willingly or accidentally training the system to give outputs that can come off as biased. Others analyzing the training of the system and hiring an ethicist may be key steps to helping make sure that the system is not biased. Otherwise, false information and discriminatory bias can end up being common when utilizing the system.

However, it may not be as simple as stated when training these systems. According to an article from Harvard Business Review, "Different cultures may also accept different definitions and ethical trade-offs—a problem for products with global markets" (Babic, 2023, Moral Risk section, para. 3). This brings up a new question regarding AI: even if a business may be using their own ethical codes to program a system, it may not follow ethical codes of other places within the world. Therefore, it is essential that a global ethical code is utilized during training, so the same ethical standards on these systems are utilized throughout the world. A key ethical code regarding AI would be UNESCO's Ethics of Artificial Intelligence code. If a company follows the guidelines set in this document and takes them into account when training their AI model, their AI model will be much less prone to certain ethical concerns such as having less biased or discriminatory answers. Having an unbiased programmer utilizing unbiased data to the best of their ability, along with having others analyze and possibly hiring an ethicist, is arguably the most vital step in business self-regulation.

The second key step within the business self regulation process is testing. It is a necessity that businesses ensure their systems are accurate before they are utilized for profit, or in ways where they have impacts on people's lives. Many private companies do not have adequate testing on their systems before they are put to use in the real world. This is especially seen within the example occurring in Broward County, where there is a key problem taking place within both of these steps. The system was clearly built with bias into it, along with it not being adequately tested before it began being utilized to essentially dictate the course of people's futures. According to MIT Technology Review, these systems are shown to be inaccurate due to testing that does not actually prove an AI system's capabilities in a specific setting. Within the article, Will Douglas Heaven states, "It's no secret that machine-learning models tuned and tweaked to near-perfect performance in the lab often fail in real settings. This is typically put down to a mismatch between the data the AI was trained and tested on and the data it encounters in the world, a problem known as data shift" (2020, para. 1). His explanation for this proves the key dilemmas going on today, where a system may be tested on specific areas related to how it was trained, but when the system encounters something in a real world setting, it ends up being inaccurate and failing. A key example of this was also described, as Heaven showed how a system used to identify diseases from clear lab pictures will end up struggling in real life, when they have to identify diseases off blurry pictures. This is just a key example of how some of the current testing on certain systems today is simply inaccurate. We simply cannot use these systems in the real world if they are not shown to be reliable with real world conditions.

Additionally, these models need to "test very extensively and test to break" (Patrick, 2019, p. 270). The possible errors within these models need to be seen beforehand so they are able to be fixed before the system is pushed out and utilized by the general public. Many businesses will simply try to get their system out into the market right away if it means more profit, but this is a major problem and defeats the purpose of self regulation. The programmed systems also need to have "an agreed and measurable accuracy rate" (Patrick, 2019, p. 270). This is a major necessity in all AI systems, within all fields. For example, many people in the

Broward County judicial system could have been saved from having longer sentences if an accuracy rate was shown for each client. Overall, testing can be a long and tedious process if done correctly. Nevertheless, it can pay off by mitigating possible biases and ethical concerns that can occur otherwise.

**Feasibility of Business Self-Regulation**

The problems with this are quite self-explanatory: will businesses actually be willing to self-regulate? After all, self-regulation all depends on if the business wants to incorporate safety and ethical codes into the company. These businesses may lose profit, but they will benefit from a moral standpoint by decreasing some of the ethical concerns that AI will cause. Companies that incorporate their systems for real-life uses will be able to have a fairer and more accurate system than if they do not undertake the process of self regulation. Less bias will be present just by the following of certain steps, but if a company's main concern is money, rather than the lives of the people its system is impacting, there will be no change in what is currently occurring. In that case, a greater solution than businesses deciding to self-regulate may just be necessary.

**A Possible Solution to These Concerns With Government Regulation**

It may be essential for there to be government regulation on AI, especially if many businesses do not impose self-regulation. Increased government regulation would be one step to help put a stop to many of the possible ethical issues discussed that occur with AI. An example of this would be by first instilling regulation to help solve the issue of discrimination from AI software. As shown in the prior paragraph about the major effects of AI bringing about discrimination, government regulation could simply block AI from being utilized within certain fields, such as job searching. Several other states throughout the United States already have possible legislation that was introduced with the aim of being implemented. For example, an article from the American Bar Association states, "The D.C. legislation, if passed, would prohibit covered entities from making an algorithmic eligibility determination on the basis of an individual's or class of individuals' actual or perceived race, color, religion, national origin, sex, gender identity or expression, sexual orientation, familial status, source of income, or disability in a manner that segregates, discriminates against, or otherwise makes important employment opportunities unavailable to an individual or class of individuals" (Wagner, 2022, What is a Bias Audit? section, para. 5). Regulations like this are ideal as they cover most of the areas that could be utilized by AI to discriminate. The regulations would allow for a fairer utilization of AI overall.

      A major issue occurring in Florida was indeed the discrimination discussed in the case of criminal prediction scores within Broward County's justice system. An article from the Center for Digital Ethics and Policy of Loyola University states how many risk assessment algorithms, such as the one utilized, are not even able to be argued against if they are inaccurate. They state, "Unfortunately, the algorithms' proprietary nature means that neither attorneys nor the general public have access to information necessary to understand or defend against these assessments" (Williams, 2018, para. 1). This emphasizes that even if these algorithms end up being incorrect on many occasions, private companies will still receive no penalty due to the general public not being able to understand how they work. This once again ties back into the issue of businesses not implementing self-regulation. A private company should not be able to keep the details of their algorithm private, especially if the algorithm is incorrect many times and has such a large impact on people's lives. Regulation should be implemented to prevent this from occurring, especially in the state of Florida. It is key for Florida state representatives to care about this

major issue and take a stand against private companies using their inaccurate systems at the expense of others.

      An act that was introduced by Congress was the Algorithmic Accountability Act of 2022, written by Senator Ron Wyden of Oregon. This act discussed how companies should know and understand the impacts that the AI systems they are using can cause (Wyden, 2022). An act such as this one truly seems like a step in the right direction, as businesses will now be forced to make more informed decisions due to understanding the major danger and disasters that could occur if AI data gets into the wrong hands. Additionally, the bill discusses how businesses will be required to have "transparency about when and how automated systems are used" (Wyden, 2022, p. 1). This will reduce the feeling of businesses thinking they can do whatever they want with user data and this new innovation in general, as they will have to explain how they utilize the AI systems and make sure they understand the risks.

      Even though this bill was not passed in 2022, it was reintroduced for 2023. Within his new summary, Senator Wyden discusses a major problem that was not addressed in the prior bill, regarding an AI system being discriminatory. He states, "A recent lawsuit alleges that a tenant screening system that disproportionately denied Black and Hispanic renter applications because it improperly relied on historical credit history and failed to consider their use of federally-funded housing vouchers" (Wyden, 2023, p. 1). This touches back on the major issue discussed prior about something needing to be done about AI being discriminatory. This bill can be a step in the right direction thanks to its inclusion of mandatory company assessments of the impacts of their systems' decisions, which is necessary in many cases such as the discrimination case occurring in Broward County discussed above.

**Feasibility of Government Regulation**

It will certainly not be easy to get new legislation passed on the topic of AI. However, there is much legislation that is already being attempted to be implemented, such as Senator Wyden's Algorithmic Accountability Act. It is absolutely necessary to support acts such as this one, as these acts will allow for a stand to be taken against many of the current AI uses that are unfortunately affecting many people's lives.

Government regulation, however, will not solve all the issues within the AI ethical crisis. Government regulation, such as the Algorithmic Accountability Act, can only do so much to deal with these ethical concerns. People's data will still be unsafe each time they use AI applications, and this is due to the fact that AI attacks just simply cannot be stopped. An article from Harvard Kennedy School shows this by stating, "Unlike traditional cybersecurity attacks, these weaknesses are not due to mistakes made by programmers or users. They are just shortcomings of the current state-of-the-art methods" (Comiter, 2019, p. 12). Therefore, these concerns cannot be completely solved. Hackers will always be present and attacks on AI are seemingly more simple just due to the methods that the AI was built with. Certain amounts of data could be under threat each time someone utilizes AI and ethical concerns will seemingly always be present.

However, implementing government regulation that can help reduce the problem is much better than sitting back and letting the problem overrun itself. Additionally, it is one key step to helping our data become safer within the AI realm. These regulations are beneficial as they can stop discriminatory AI uses and create regulation on what businesses are allowed to utilize AI and its specific data for.

**Conclusion**

      There is no quick or simple solution to many of the ethical challenges faced today due to AI. However, government regulation and business self-regulation are some key steps that could help on our path to limit many of artificial intelligence's current ethical concerns. A possible mix of these two forms of regulation may just prove best, but it is essential for at least one form of regulation to be put into place for any change to occur.